\documentstyle[emulateapj]{article}

\catcode`\@=11 
\def\@versim#1#2{\vcenter{\offinterlineskip
        \ialign{$\m@th#1\hfil##\hfil$\crcr#2\crcr\sim\crcr } }}

\newcommand{\beq}{\begin{equation}}
\newcommand{\eeq}{\end{equation}}
\def\lsim{\mathrel{\mathpalette\@versim<}}
\def\gsim{\mathrel{\mathpalette\@versim>}}

\def\mo{\dot m_{o}}
\def\ro{r_{o}}

\def\mpy{M_\odot \ {\rm yr^{-1}}}

\begin{document}

\title{What is the Accretion Rate in Sgr A$^*$?}  
\author{Eliot Quataert, Ramesh Narayan, and Mark J. Reid}
\affil{Harvard-Smithsonian Center for Astrophysics, 60 Garden St.,
  Cambridge, MA 02138; equataert, rnarayan, mreid @cfa.harvard.edu}

\medskip

\begin{abstract}
  
  The radio source Sgr A$^*$ at the center of our Galaxy is believed
  to be a $2.6 \times 10^6 M_\odot$ black hole which accretes gas from
  the winds of nearby stars.  We show that limits on the X-ray and
  infrared emission from the Galactic Center provide an upper limit of
  $\sim 8\times 10^{-5} \mpy$ on the mass accretion rate in Sgr A$^*$.
  The advection-dominated accretion flow (ADAF) model favors a rate
  $\lsim 10^{-5}\mpy$.  In comparison, the Bondi accretion rate onto
  Sgr A$^*$, estimated using the observed spatial distribution of mass
  losing stars and assuming non-interacting stellar winds, is $\sim 3
  \times 10^{-5} \mpy$. There is thus rough agreement between the
  Bondi, the ADAF, and the X-ray inferred accretion rates for Sgr A*.
  We discuss uncertainties in these estimates, emphasizing the
  importance of upcoming observations by the Chandra X-ray observatory
  (CXO) for tightening the X-ray derived limits.

\ 

\noindent {\em Subject Headings:} accretion, accretion disks -- 
black holes -- Galaxy: center -- radio sources: Sgr A*

\end{abstract}
 
\section{Introduction}

Stellar kinematics show that there are $\approx 2.6 \times 10^6
M_\odot$ within $\approx 0.015$ pc of the Galactic Center (Eckart \&
Genzel 1997, Ghez et al. 1998), centered on the radio source Sgr A*
(Menten et al. 1997).  The most plausible explanation is that Sgr
A$^*$ is a $\approx 2.6 \times 10^6 M_\odot$ accreting black hole.

Sgr A$^*$ is believed to accrete the winds from nearby ($\sim 0.1$ pc)
massive stars (Krabbe et al. 1991).  Hydrodynamical simulations of the
Galactic Center region, assuming 10 randomly distributed point sources
(to mimic the nearby stars) and a $10^6 M_\odot$ black hole, yield
accretion rates $\approx 1-2 \times 10^{-4} \mpy$ (Coker \& Melia
1997; hereafter CM).  If we naively increase this by a (Bondi capture)
factor of $2.6^2$ to account for the actual mass of the central
object, the theoretically predicted accretion rate becomes $\sim
10^{-3} \mpy$ (but see \S3).

The bolometric luminosity of Sgr A$^*$ is observed to be $\lsim
10^{37}$ ergs s$^{-1}$ (Genzel et al. 1994, Narayan et al. 1998).  For
an accretion rate of $\sim 10^{-3} \mpy$, this corresponds to a
radiative efficiency of $\sim 10^{-7}$!  A possible explanation for
the low luminosity of Sgr A$^*$ is that the accretion occurs via an
advection-dominated accretion flow (ADAF; Narayan, Yi \& Mahadevan
1995, Manmoto et al. 1997, Narayan et al. 1998; see, e.g., Melia 1992,
1994, Mastichiadis \& Ozernoy 1994, Falcke 1996, Beckert \& Duschl
1997 for alternative models of Sgr A$^*$).  All ADAF models in the
literature, however, require an accretion rate of $\lsim 10^{-5} \mpy$
(see Table 2 of Quataert \& Narayan 1999, which gives accretion rates
in Eddington units; their estimates need to be multiplied by a factor
of 0.055 for $\mpy$).  If CM's black hole mass scaled Bondi capture
estimate of $\sim 10^{-3} \mpy$ is taken at face value, there is a
discrepancy of a factor of $\gsim 10^2$ in mass accretion rates.

In \S2, we sharpen this discrepancy by using spectral observations of
the Galactic Center to constrain the accretion rate of gas at large
distances from the central black hole.  The argument is nearly
independent of the accretion model employed.  We then calculate Bondi
capture estimates of accretion onto Sgr A$^*$, explicitly using the
observed spatial distribution and wind properties of stars in the
Galactic Center (\S3).  In \S4 \& \S5 we conclude and assess the
agreement between various theoretical estimates of the accretion rate
in Sgr A$^*$.

\section{Observational Constraints on $\dot M$}

We use X-ray and IR limits on the emission from the Galactic Center to
derive constraints on the accretion rate of optically thin or thick
gas onto Sgr A$^*$.  We take the distance to the Galactic Center to be
$8.0$ kpc.  In what follows, $m_{S}$ denotes the mass of the black
hole in units of $2.6 \times 10^6 M_\odot$, $R$ denotes the physical
radius in the accretion flow, $r$ ($=R/7.38\times10^{11}m_S$ cm)
denotes the radius in Schwarzschild units, and $\dot m$ is the
accretion rate in Eddington units, where $\dot M_{edd} = 0.055 m_{S}
M_\odot$ yr$^{-1}$.

\subsection{X-ray Constraints on Optically Thin Accretion}

If the gas accreting onto Sgr A$^*$ is optically thin, it will emit
X-rays by bremsstrahlung emission. The observed limits on X-ray
emission from Sgr A$^*$ can therefore be used to constrain the
accretion rate.  The accretion rate of the gas is related to the
density, $\rho$, by the continuity equation \beq \dot M = 4 \pi R^2
\rho v_{ff} \eta_H \eta_v = \dot M_o \left(r/r_o \right)^p,
\label{cont} \eeq  where $r_o \sim 10^5$, the ``outer'' radius of the flow, is 
roughly the Bondi capture radius of the black hole. The radial
velocity of the gas is related to the free fall velocity, $v_{ff}$, by
$|v| = \eta_v v_{ff}$ and the vertical scale height of the gas is $H =
\eta_H R$.  For a pure Bondi flow, $\eta_v = \eta_H = 1$, while
for an ADAF, $\eta_v \approx \alpha$ (the viscosity parameter) and
$\eta_H \approx 1/2$.  

Equation (\ref{cont}) allows for the possibility that the mass
accretion rate may decrease with radius due to an outflow/wind (in
which case the assumed geometry is a ``thick'' equatorial inflow with
a bipolar outflowing wind; cf Blandford \& Begelman 1999).  The
accretion rate at the outer edge of the flow (i.e., the mass supply
rate) is $\dot M_o$ and the parameter $p$ measures the strength of the
outflow.

We parameterize the electron temperature in the accretion flow by $T_e
= \eta_T T_o (r/r_o)^{-a}$, where $T_o = 10^{12}/r_o$ is $\sim$ the
proton virial temperature at $r_o$ and $a$ ($\sim 1$ at large radii)
determines the radial profile of the electron temperature; $\eta_T
\approx 1$ for both Bondi accretion and an ADAF.

The bremsstrahlung luminosity at frequency $\nu$ is given by \beq
L_\nu = 4 \pi \eta_H \int \epsilon_\nu R^2 dR, \label{brem} \eeq where
$\epsilon_\nu = 2.6 \times 10^{10} \rho^2 T_e^{-1/2} \exp[-h \nu/k
T_e]$ ergs s$^{-1}$ cm$^{-3}$ is the bremsstrahlung emissivity (e.g.,
Rybicki \& Lightman 1979; we have ignored a weak frequency-dependent
Gaunt factor).  Carrying out the integral in equation (\ref{brem})
using the above expressions for the density and temperature of the
gas, and approximating the exponential cutoff in $\epsilon_\nu$ by a
step function, we find (in ${\rm ergs \ s^{-1} \ Hz^{-1}}$) \beq L_\nu
\approx { 6 \times 10^{19} \over 2p + a/2} \eta_H^{-1} \eta_v^{-2}
\eta_T^{-1/2} \mo^2 m_S \ro^{1/2} \left(r_\nu \over r_o\right)^{2p +
  a/2},
\label{lum} \eeq where $\mo = \dot M_o/\dot M_{edd}$ and 
\beq r_\nu = r_o \ {\rm Min}\left[1, \left( { h \nu \over \eta_T k
      T_o} \right)^{-1/a} \right].
\label{rnu} \eeq

Observations at frequency $\nu$ are dominated by emission from the
radius $r_\nu$.  In particular, if the X-ray energies of interest are
less than the {\em minimum} electron thermal energy in the flow (which
is obtained at $\ro$ and is $\sim 1$ keV for $r_o \sim 10^5$), the
bremsstrahlung luminosity is dominated by emission from the largest
radii in the flow.  In this case $r_\nu = r_o$ and the parameters $p$
and $a$, which determine the radial profiles of the density and
electron temperature, are irrelevant; they enter only in the
normalization of the emission, not in the exponent of any parameters
of the problem.  As a result, observations of soft X-rays provide
effective constraints on $\dot M_o$, the rate at which matter is
supplied to the accretion flow on the outside.  They cannot, however,
constrain the flow structure closer to the black hole (e.g., the
nature or strength of an outflow).

Parameterizing the observational constraints on emission from the
Galactic Center via $\nu L_\nu \lsim 10^{34} L_{34}$ ergs s$^{-1}$ at
$h \nu = \nu_1$ keV ($\nu = 2.4 \times 10^{17} \nu_1$ Hz) equation
(\ref{lum}) can be inverted to give a constraint on the accretion rate
of the gas (taking $2p + a/2 \approx 1$ in the denominator of eq.
[\ref{lum}] for simplicity):
\begin{eqnarray} \label{mdot} \dot M_o &\lsim& 8 \times  10^{-5} r_5^{-1/4} 
\eta_H^{1/2} \eta_v \eta_T^{1/4}
  m_S^{1/2} \\ \nonumber & \times & \left(L_{34} \over
    \nu_1\right)^{1/2} \left( r_o \over r_\nu\right)^{p + a/4} \ {\rm
    M_\odot \ yr^{-1}},  \end{eqnarray} where $r_5 = r_o/10^5$.

The parameters $\eta_v, \eta_H,$ and $\eta_T$ must all be $< 1$.
Furthermore, the Bondi capture radius in Sgr A$^*$ for wind material
moving at $\sim 10^3$ km s$^{-1}$ is $\sim 10^5$ Schwarzschild radii
so that the accretion flow must originate at $r_5 \sim 1$.

ROSAT observations of the Galactic Center indicate that the X-ray
luminosity of Sgr A$^*$ at $\approx 1$ keV is $\lsim 10^{34}$ ergs
s$^{-1}$, in which case $L_{34} \approx \nu_1 \approx 1$ and $r_\nu
\approx r_o$ (but, see Narayan et al. 1998 for a discussion of the
effect of uncertainties in the absorbing column; we take $N_H = 6
\times 10^{22}$ cm$^{-2}$).  By equation (\ref{mdot}) this implies
$\dot M_o \lsim 8\times10^{-5} {\rm M_\odot \ yr^{-1}}$; this is a
strict upper limit on the accretion rate of optically thin gas in Sgr
A$^*$.


ASCA observations of the Galactic Center (Koyama et al. 1996) give a
$2-10$ keV limit of $10^{35}$ ergs s$^{-1}$.  Taking $\nu_1 \approx
5$, equation (\ref{mdot}) gives a limit of $\dot M_\nu \lsim 10^{-4}
\mpy$ at $r_\nu \approx 0.2 r_o \sim 10^4$, where $\dot M_\nu = \dot
M_o (r_\nu/r_o)^p$ is the accretion rate at the radius $r_\nu \approx
r_o \nu_1^{-1/a} \eta_T^{1/a}$.  SIGMA observations (Vargas et al.
1998) give $\nu L_\nu \lsim 2 \times 10^{35}$ ergs s$^{-1}$ at $\nu_1
\approx 75$, which requires $\dot M_\nu \lsim 10^{-4} \mpy$ at $r_\nu
\approx 0.01 r_o \sim 10^3$.  If there are no winds emanating from Sgr
A* at $r \sim r_0$ (i.e., $p \approx 0$) then the ASCA and SIGMA
observations require $\dot M_o \lsim 10^{-4} \mpy$.  This may be an
even stronger limit than that derived from the ROSAT observations,
since it is insensitive to the absorbing column.

\subsection{Infrared Constraints on Optically Thick Accretion}

The preceding subsection assumes that the gas is optically thin at
large radii.  We believe that this is the most likely scenario.  It is
worth, however, examining observational constraints imposed on the
accretion rate of an optically thick disk.

For a standard Shakura-Sunyaev disk (e.g., Frank, King, \& Raine
1992), the luminosity is given by \beq \nu L_\nu = {16 \pi^2 h \cos(i)
  \nu^4 \over c^2} \int^{R_{out}}_{R_{in}} {R dR \over \exp(h
  \nu/kT(R)) - 1}, \label{disk} \eeq where $i$ is the inclination of
the disk, $R_{out}$ and $R_{in}$ are the inner and outer radii of the
disk, and $T(R)$, the effective temperature of the disk, depends on
the accretion rate (cf Frank et al. 1992).

Menten et al. (1997) have obtained an upper limit of $\approx 10^{35}$
ergs s$^{-1}$ on the 2.2 $\mu m$ emission from the Galactic Center.
This emission strongly constrains the properties of any geometrically
thin disk.  It is straightforward to numerically integrate equation
(\ref{disk}) and obtain the 2.2 $\mu$ emission from the disk.
Requiring this to be $\lsim 10^{35}$ ergs s$^{-1}$ yields a constraint
on the parameters $\dot M$ and $r_{in}$.  For $i \sim 60^o$, taking
$\dot M \approx 10^{-3} {\rm M_\odot \ yr^{-1}}$ implies $r_{in} \gsim
10^4$.  If $\dot M \approx 10^{-5} {\rm M_\odot \ yr^{-1}}$ the
constraint is slightly less stringent, $r_{in} \gsim 3 \times 10^3$
(note that, to satisfy all existing IR and radio limits on emission
from Sgr A* typically requires {\em even larger} values of
$r_{in}$).\footnote{Genzel et al. (1997) report a possible detection
  of Sgr A* at 2.2 $\mu m$, at a flux level above Menten et al.'s
  upper limit (indicating possible variability in the source).  This
  need not represent a detection of an optically thick disk, but could
  be either (1) singly Compton scattered synchrotron emission, (2)
  synchrotron emission from non-thermal electrons, or (3) optically
  thick accretion disk emission.  {\em If} interpreted as the latter,
  the upper limits in this paragraph become (approximate) equalities.}

This implies that, for any reasonable inclination and accretion rate,
a thin disk cannot extend much inside $r_{in} \sim 10^4$ without
violating the IR limits on emission from the Galactic center.  Even if
we assume that from the Bondi capture radius to $r_{in}$ the accretion
is via a geometrically thin disk, and then for $r < r_{in}$ it is
optically thin, our limits on $\dot M$ from the previous subsection
are not changed substantially.  This is because the X-ray constraints
on the accretion rate are quite insensitive to the precise outer
boundary of the optically thin region (cf. eq.  [\ref{mdot}] where the
limit is $\propto \ro^{-1/4}$).


\section{Bondi Capture Estimates}

There are $\approx10$ He I stars (Krabbe et al. 1991) projected within
$10$ arcsec of the position of Sgr A$^*$ (Menten et al.  1997).  The
winds from these stars result in mass loss rates ranging from about
$8\times10^{-5}$ to $8\times10^{-4}$ $\mpy$ and might provide most of
the matter accreted by Sgr A$^*$.  

Using the distribution of the He I stars projected on the plane of the
sky, their mass loss rates, $\dot M_w$, and their wind speeds, $V_w$
(Najarro et al. 1997), we have estimated the mass accretion rate onto
a $2.6\times10^6 M_\odot$ black hole.  The line-of-sight
(z-coordinate) offset of each star relative to Sgr A$^*$ (the black
hole) is unknown, and we assigned a z-distance randomly to each star,
assuming a Gaussian distribution.  The stellar wind material passing
within a radius $R_A=2GM_{BH}/V_w^2$ of Sgr A$^*$ was assumed to be
accreted.  This resulted in a contribution to the accretion rate for a
given star, $\dot M_{acc}$, given by \beq \dot M_{acc} = {1\over2}
{\dot M_w} \left( 1 - \sqrt{ 1 - R_A^2/r^2} \right),
\label{acc} \eeq where $r$ is the three-dimensional distance of the
star from Sgr A$^*$ (Note that all stars in the Najarro et al. sample
have $r > R_A$).

Using equation (\ref{acc}), a star such as IRS~16C with $\dot M_w
\approx 8\times10^{-5}$ $\mpy$, a wind speed of 650 $\rm km \ s^{-1}$,
and a true distance from Sgr A$^*$ of $\approx2$~arcsec (its projected
distance is 1.4~arcsec) would contribute about $1.4\times10^{-5}$
$\mpy$ to the total mass accretion rate of Sgr A$^*$.  On the other
hand, the most powerful He I star near Sgr A$^*$, IRS~13E1, which
loses mass at a rate of $8\times10^{-4}$ $\mpy$ with a wind speed of
$1000 \rm \ km \ s^{-1}$, has a projected distance from Sgr A$^*$ of
3.7 arcsec and thus makes a {\it maximum} contribution to the mass
accretion rate of only $5\times10^{-6}$ $\mpy$.

For each trial distribution of stars, obtained for different
realizations of the random z-coordinate distances from Sgr A$^*$, we
calculated the total mass accretion rate as the sum of the accretion
rates for each star.  This assumes no wind--wind interactions.
Varying the standard deviation of the z-coordinate distribution,
$\sigma_z$, we found average total mass accretion rates of $\dot M_o =
5, 3,$ and $2 \times 10^{-5}$ $\mpy$ for $\sigma_z$ of 2, 5, and 8
arcseconds, respectively.\footnote{For an ``average'' wind speed of
  $700 v_7$ km s$^{-1}$, this corresponds to $\sigma_z$ of $2,\ 5, \ 
  \& \ 8 \ v^{-2}_7$ $R_A$.}  If the He I stars have a $\sigma_z
\approx 5$, as suggested by the projected spatial distribution of
stars near Sgr A$^*$ (Genzel et al.  1997; Ghez et al. 1998), then the
total mass accretion rate for non-interacting stellar winds onto a
$2.6\times10^6$ $M_\odot$ black hole should be $\dot M_o \approx
3\times10^{-5}$ $\mpy$.

This accretion rate is noticeably smaller than the value of $\sim
10^{-3} \mpy$ obtained by scaling CM's numerical results from a
$10^6M_\odot$ black hole to a $2.6\times10^6$ $M_\odot$ black hole
(cf. \S1).  The main reason for the discrepancy seems to be that the
spatial distribution of point sources they take (e.g., $\pm 3 R_A$) is
a reasonable approximation of the observed stellar distribution only
for a $2.6\times10^6 M_\odot$ black hole, not a $10^6M_\odot$ hole.
That is, for $v_w \approx 700$ km s$^{-1}$ and a black hole mass of
$10^6 M_\odot$, $3 R_A$ is only $\approx 1''$, which is smaller than
the typical projected distance of the mass losing stars from Sgr A*
(on the other hand, for a $2.6 \times 10^6 M_\odot$ black hole, $3
R_A$ corresponds to $\approx 3''$).  One should therefore not scale
CM's result up by a factor of $2.6^2$.  This leaves a smaller residual
discrepancy, perhaps due to wind-wind interactions, which they treat
in their hydrodynamical simulations, but which we neglect.

\section{Discussion}

In \S2 of this paper, we showed that ROSAT X-ray observations of Sgr
A$^*$ provide a firm limit on the accretion rate at large radii,
namely $\dot M_o \lsim 8 \times 10^{-5} \eta_v L_{34}^{1/2} \mpy$ at
$r \approx 10^5$, where $\eta_v$ is the ratio of the radial velocity
of the gas to the free fall velocity and $L_{34}$ is the soft X-ray
($\approx 1$ keV) luminosity of the source.  We obtain similar limits
at $r \sim 10^4$ and $r \sim 10^3$ using ASCA and SIGMA data,
respectively.  It is important to emphasize that these limits
correspond to the X-ray radiation being dominated by bremsstrahlung
emission.  If part of the X-rays are produced by Comptonization, as in
some models, then the accretion rates must be even lower than the
limits derived here.

The X-ray luminosity of $10^{34}~{\rm erg\,s^{-1}}$ used in the above
scalings is itself an upper limit.  There is considerable diffuse
emission from the Galactic Center (Koyama et al. 1996), and the
luminosity of the accretion flow is uncertain.  Another uncertainty is
the absorbing column, which could lie anywhere from the value used
here, $6\times10^{22}~{\rm cm^{-2}}$ (Watson et al. 1981), to
$1.5\times10^{23}~{\rm cm^{-2}}$ (Predehl \& Trumper 1994).  If the
column is closer to the higher value, then the unabsorbed luminosity
of Sgr A* could be larger than we have assumed.  High angular
resolution observations with CXO (formerly AXAF) should solve both
problems.

Sgr A* probably accretes most of its mass from the winds of nearby
massive stars (Krabbe et al.  1991).  We have estimated the Bondi
accretion rate onto Sgr A$^*$ using the observed spatial distribution
of these stars, along with their estimated mass loss rates and outflow
velocities (\S3).  The resulting accretion rate is $\dot M_o \approx 3
\times 10^{-5} \mpy$ for plausible values of the unknown spatial
distribution of stars along the line of sight.  This estimate of $\dot
M_o$ should perhaps be interpreted as a lower limit, as there may be
other sources of matter for the accretion flow, and wind-wind
interactions and post-shock cooling may increase the fraction of the
wind material accreted.  

The upper limit on $\dot M_o$ from the X-ray observations and the
lower limit on $\dot M_o$ from the capture of stellar winds are
roughly compatible, requiring $\eta_v \sim 1$ for the accretion flow
and $L_{34} \sim 1$ for the (unknown) X-ray luminosity.  If the
accretion onto Sgr A$^*$ occurs viscously via an ADAF, then $\eta_v
\approx \alpha$.  For values of $\alpha \approx 0.01$ seen in
numerical simulations of thin disks (Hawley, Gammie, \& Balbus, 1996),
there is a substantial discrepancy between the Bondi and X-ray
inferred accretion rates, while for the values of $\alpha \approx
0.1-0.3$ favored by Narayan et al.  (1998; see also Esin et al.
1997), the two estimates are comparable.

Blandford \& Begelman (1999) have recently argued that only a small
fraction of the mass supplied to an ADAF reaches the central black
hole, with most of it being lost to an outflow/wind.  It is important
to emphasize that this alone will not modify the conclusions of this
paper.  Our primary constraint on the accretion rate utilizes the
$\sim 1$ keV X-ray flux, which is dominated by bremsstrahlung emission
from large radii in the flow ($r \approx r_0 \approx 10^5$); at such
radii, any wind from the accretion flow has little effect.

Gruzinov (1999) has argued that radial heat conduction would
significantly reduce the mass accretion rate in Bondi flows.  This
explanation is also unlikely to effect the results of this paper.  The
density at $\sim r_0$ is a boundary condition set by the stellar
winds.  Gruzinov's flow would therefore produce the same X-ray
luminosity from radii $r \sim r_0$ as the models we have considered
(at large radii $\sim r_0$, Gruzinov's reduced accretion rate is due
to a smaller radial velocity, not a lower density).

Finally, it is worth emphasizing that our paper does not explicitly
address the reason for the low radiative efficiency of Sgr A*.  This
is determined by the physical conditions close to the black hole ($r
\lsim 10-100$), while our analysis focuses on larger radii.

\section{Conclusion}

With current data, we believe that there is rough agreement between
the Bondi accretion rate (as estimated in this paper), the limits
inferred from X-ray and IR observations, and the value (or, more
precisely, the upper limit) favored by ADAF models. If this
straightforward interpretation is correct, CXO should detect an X-ray
flux comparable to that of ROSAT, despite its significantly improved
angular resolution.  On the other hand, if CXO substantially lowers
the soft X-ray flux from the Galactic Center there would be an
inconsistency between the Bondi rate and the values favored by the
X-ray observations.  Given the uncertainties in the accretion rate
estimates, it is worth spelling out what we believe to be the two
plausible, if somewhat mundane, explanations if such an inconsistency
is discovered.

The first assumes that the Bondi accretion rate estimates are correct
and $\dot M_o \gsim 3 \times 10^{-5} \mpy $.  This could be reconciled
with the limits inferred from a significantly smaller soft X-ray flux
only if $N_H$ is larger, by a factor of few, than the value we have
assumed ($ 6 \times 10^{22}$ cm$^{-2}$). 

The other alternative is simply that the Bondi accretion rate
estimates at large radii are in error, with the true accretion rate
being $\lsim 10^{-5} \mpy$.  We see several possible reasons for such
an error:

1. The estimates of the stellar mass loss parameters, $\dot M_w$ and
$V_w$, may be in error. Note that a $25 \%$ uncertainty in $V_w$
translates to a factor of $2.5$ uncertainty in the Bondi accretion
rate.  Together with a factor of few uncertainty in the stellar mass
loss rates, $\dot M_w$, the Bondi accretion rate is probably uncertain
by at least a factor of $\sim 5$.

2. If the stars in the Galactic Center are not randomly distributed
around Sgr A*, but instead have a large z-coordinate offset, the
predicted accretion rate will be reduced.  We find that a systematic
offset of $\gsim 0.5$ pc along the line of sight is needed to bring
$\dot M_o \lsim 10^{-5} \mpy$.

3. Perhaps Sgr A$^*$ has a strong outward wind which maintains a
low-density ``bubble'' around the source (the winds from the stars in
the small stellar cluster within 1'' of Sgr A* -- possibly 09 stars;
Genzel et al. 1997 -- might also contribute sufficient momentum to
partially impede accretion of the HeI stars winds).  If this bubble
extends beyond the accretion radius $R_A$, then the accretion rate
could be significantly reduced.  The bubble is likely to be held up by
mechanical pressure from a wind rather than radiative heating since
the luminosity of Sgr A$^*$ is much too low for the latter to be
efficient.

\acknowledgements We thank Rob Coker for useful conversations and the
referee for useful comments. We acknowledge support from an NSF
Graduate Research Fellowship (E.Q.)  and NAG 5-2837 (R.N.).


\begin{thebibliography}{999}
\bibitem{}Beckert, T., \& Duschl, W. J. 1997, A\&A, 328, 95 
\bibitem{}Blandford, R. D. \& Begelman, M. C., 1999, MNRAS in press (astro-ph/9809083)
\bibitem{} Coker, R. \& Melia, F., 1997, ApJ, 488, L14 (CM)
\bibitem{} Eckart, A., \& Genzel, R. 1997, MNRAS, 284, 576 
\bibitem{} Esin, A. A., McClintock, J. E., \& Narayan, R., 1997, ApJ, 489, 86
\bibitem{}Falcke, H., 1996, in IAU Symp. 169, Unsolved Problems of the Milky Way, eds. L. Blitz \& P. J. Teuben (Dordrecht: Kluwer), 163
\bibitem{}Frank, J., King, A., \& Raine, D. 1992, Accretion Power in Astrophysics, (Cambridge: Cambridge Univ. Press
\bibitem{} Genzel, R., Eckart, A., Ott, T., \& Eisenhauer, F., 1997, MNRAS, 291, 219
\bibitem{} Genzel, R., Hollenbach, D., \& Townes, C. H. 1994, Rep Prog. Phys., 57, 417
\bibitem{} Ghez, A. M., Klein, B. L., Morris, M., \& Becklin, E. E., 1998, ApJ, 509, 678
\bibitem{}Gruzinov, A., 1999, ApJ in press (astro-ph/9809265)
\bibitem{} Hawley, J. F., Gammie, C. F., \& Balbus, S. A., 1996, ApJ, 464, 690 
\bibitem{}Koyama, K. et al. 1996, PASJ, 48, 249
\bibitem{} Krabbe, A., Genzel, R., Drapatz, S. \& Rotaciuc, V., 1991, ApJ, 382, L19
\bibitem{}Manmoto, T., Mineshige, S., \& Kusunose, M., 1997, ApJ, 489, 791
\bibitem{}Mastichiadis, A., \& Ozernoy, L. M., 1994, ApJ, 426, 599--603
\bibitem{}Melia, F., 1992, ApJ, 387, L25
\bibitem{}Melia, F., 1994, ApJ, 426, 577
\bibitem{} Menten, K. M., Reid, M. J., Eckart, A., \& Genzel, R. 1997, ApJ, 475, L111
\bibitem{}Najarro, F. et al. 1997, A \& A, 325, 700
\bibitem{} Narayan, R., Mahadevan, R., Grindlay, J.E., Popham, R.G., \& Gammie, C., 1998a, ApJ, 492, 55
\bibitem{}Narayan, R., Yi, I., \& Mahadevan, R., 1995, Nature, 374, 623
\bibitem{}Predehl, P., \& Trumper, J. 1994, A\&A, 290, L29
\bibitem{}Quataert, E. \& Narayan, R., 1999, ApJ in press  (astro-ph/9810136)
\bibitem{}Rybicki, G. \& Lightman, A., 1979, { \em Radiative Processes in Astrophysics} (New York:  John Wiley \& Sons
\bibitem{}Vargas. M. et al., 1996, in {\em The Galactic Center}, ed. R. Gredel, 431 
\bibitem{}Watson, M. G., Willingale, R., Grindlay, J. E., \& Hertz, P., 1981, ApJ, 250, 142
\bibitem{}

\end{thebibliography}
\end{document}